\documentclass[preprint2]{aastex}
\usepackage{xspace} 
\usepackage[free-standing-units]{siunitx} 

\newcommand{\ww}{WASP-39b\xspace}
\newcommand{\wq}{WASP-43b\xspace}

\shorttitle{\ww and \wq transit observations}
\shortauthors{Ricci et al.}

\begin{document}


\title{Multi-filter transit observations of \ww and \wq 
  \\ with three San Pedro M\'artir telescopes}


\author{D.~Ricci, 
  F.~G.~Ram\'on-Fox, 
  C.~Ayala-Loera, 
  R.~Michel, 
  S.~Navarro-Meza, 
  L.~Fox-Machado, and 
  M.~Reyes-Ruiz 
}

\affil{Observatorio Astron\'omico Nacional, Instituto de Astronom\'ia --
  Universidad Nacional Aut\'onoma de M\'exico, Ap. P. 877, Ensenada, BC
  22860, Mexico}\email{indy@astrosen.unam.mx}

\and
\author{S.~Brown~Sevilla}
\affil{ Facultad de Ciencias F\'isico-Matem\'aticas,
  Benem\'erita Universidad Aut\'onoma de Puebla,
  Av. San Claudio y 18 Sur, 72570 Puebla, Mexico.}

\and
\author{S.~Curiel}
\affil{Instituto de Astronom\'ia -- Universidad Nacional Aut\'onoma de
  M\'exico, Ap. P. 70-264, M\'exico, D.F. 04510, Mexico}


\begin{abstract}
  Three optical telescopes located at the San Pedro M\'artir National
  Observatory were used for the first time to obtain multi-filter
  defocused photometry of the transiting extrasolar planets \ww and
  \wq.  We observed \ww with the $2.12\meter$ telescope in the $U$
  filter for the first time, and additional observations were carried
  out in the $R$ and $I$ filters using the $0.84\meter$ telescope. \wq
  was observed in $VRI$ with the same instrument, and in the $i$
  filter with the robotic $1.50\meter$ telescope.  We reduced the data
  using different pipelines and performed aperture photometry with the
  help of custom routines, in order to obtain the light curves.  The
  fit of the light curves ($1.5$--$2.5\milli\rm mag$ rms), and of the
  period analysis, allowed a revision of the orbital and physical
  parameters, revealing for \ww a period
  ($4.0552947\pm9.65\times10^{-7}$ days) which is $3.084\pm0.774$
  seconds larger than previously reported. Moreover, we find for \wq a
  planet/star radius ($0.1738\pm0.0033$) which is $0.01637\pm0.00371$
  larger in the $i$ filter with respect to previous works, and
    that should be confirmed with additional observations.  Finally,
  we confirm no evidence of constant period variations in \wq.
\end{abstract}


\section{Introduction}
\label{info}

Since the first detection of an exoplanet orbiting around a pulsar
\citep{1992Natur.355..145W,1994Sci...264..538W} and around a main
sequence star \citep{1995Natur.378..355M} using radial velocity
measurements, several other techniques, such as microlensing,
coronagraphy, and polarimetric observations have been developed for
the search of other worlds. Transit observations have contributed to a
large fraction of the about $1\,800$ confirmed
planets\footnote{\url{http://exoplanetarchive.ipac.caltech.edu/cgi-bin/ExoTables/nph-exotbls?dataset=planets}}.
Several systematic surveys have been implemented for the detection of
exoplanet candidates by transit observations.

Important examples of these surveys include space-based missions such
as CoRoT \citep{2009IAUS..253...71B} and Kepler
\citep{2010Sci...327..977B} satellites. The high photometric precision
of their observations allowed to boost the number of candidates and
confirmed extrasolar planets detected in the last five years.  The
next generation of space-based missions, such as the orbiting GAIA
\citep{2009sf2a.conf...27B} and the upcoming JWST \citep{jwst, jwst2},
PLATO \citep{plato}, and TESS \citep{tess} are expected to increase
the detection rate, and broaden the limits of the currently observed
distribution of exoplanetary parameters, despite that some of these
missions have not the detection of exoplanets as a main goal.

On the other hand, ground-based photometric surveys such as SuperWASP
\citep{2006PASP..118.1407P} and HAT \citep{2007ApJ...656..552B} are
focused on exoplanets with large radii and orbiting close to their
host star, as these objects are particularly suited for these
instruments.  In order to improve the photometric precision of
ground-based follow-up observations of exoplanetary transits,
a technique consisting in slightly defocusing the image has been
introduced and widely used in recent works \citep{2009MNRAS.396.1023S,
  2009MNRAS.399..287S, 2010MNRAS.408.1680S, 2012MNRAS.426.1338S,
  2013MNRAS.434.1300S, 2014MNRAS.444..776S}. The achieved results
often compete with those of space-based surveys, and increase the
opportunity for small telescopes to produce light curves with a
remarkable photometric precision.

Following the results obtained with the defocused photometry
technique, we started an observational campaign of known exoplanetary
transits, already detected by the SuperWASP and HAT surveys. 
Photometric follow-up of these objects in different bandpasses is
useful to study a possible dependence of the planetary radius as a
function of the wavelength.

This project makes use of all three telescopes operating at the
San Pedro M\'artir National Astronomical Observatory (see
Sect.~\ref{obs} for details), which we employ for the first time
in the field of transit observations.  In this paper we focus on two
particular targets: \object{WASP-39b} and \object{WASP-43b}, which are
briefly presented in the following subsections.

\begin{table*}[t]
  \caption{Features of the three San Pedro M\'artir 
    telescopes and CCD detectors used for the observation
    of \ww and \wq. 
\label{tab:tel}}
\centering
\begin{tabular}{llccccl}
\tableline
\tableline
  Diameter   & focal ratio &      CCD size       &    Resolution    &           Field of View           &  Read Out Noise  & Gain   \\
\tableline                                                                           
$0.84\meter$ &  $f/15.0$   & $2043\times4612$ px & $0.25\arcsec/$px &   $8.4\arcmin\times19.0\arcmin$   & $\;\;4.8\rm e^-$ & 1.8 \\
$1.50\meter$ &  $f/13.0$   & $1040\times1056$ px & $0.32\arcsec/$px & $5.6\arcmin\times\;\; 5.6\arcmin$ &  $11.3\rm e^-$   & 4.0 \tablenotemark{a} \\
$2.12\meter$ &  $f/13.5$   & $2048\times2048$ px & $0.18\arcsec/$px & $6.2\arcmin\times\;\; 6.2\arcmin$ & $\;\;4.5\rm e^-$ & 1.8   \\
\tableline
\end{tabular}
 \tablenotetext{a}{binning $2\times2$.}
\end{table*}

\subsection{\ww}

\ww is a Saturn-mass planet recently discovered by the SuperWASP
survey, and confirmed by \cite{w39} with a photometric and
spectroscopic follow-up observations.
The photometry was carried on with the Pan-STARRS-$z$ filter on the
the $2\meter$ Faulkes telescope North, and in the Gunn-$i$ filter on
the $1.2\meter$ Euler telescope.  The spectroscopy was carried out
with the SOPHIE \citep{2008SPIE.7014E..17P} and the CORALIE
\citep{1996A&AS..119..373B, 2000fepc.conf..548Q} spectrographs,
installed on the $1.93\meter$ telescope at the \emph{Haute-Provence}
Observatory (OHP) and on the $1.2\meter$ Euler telescope,
respectively.

\cite{w39} obtain a period of $P=4.055259$ days, a $T_0 =
2\,455\,342.9688$ in Heliocentric Julian Date (HJD), a transit
duration of $168.192$ minutes, and the following physical parameters:
a mass of $0.28 M_J$, a radius of $1.27 R_J$, and thus a resulting
mean density of $0.14 \rho_J$. This is one of the lowest values for
the currently known extrasolar planets.  The reported semi-major axis
is $0.0486$ AU, and evidences of a highly inflated radius are also
found.

This work presents the first photometric follow-up of \ww since the
discovery.

\subsection{\wq}

\wq is a very short-period hot Jupiter-like planet discovered by
\cite{w43} using observations from the WASP-South camera array,
complemented with photometric observations with the $0.6\meter$
TRAPPIST telescope in the $I+z$ band and the Euler telescope in the
Gunn $r$ band. Radial velocity measurements were also obtained with
CORALIE.
\cite{w43} find a period of $P=0.813475$ days, an initial epoch
of $T_0 = 2\,455\,528.86774$ HJD, and a duration of $69.552$ min.  The
authors derive a mass of $1.8 M_J$, a radius of $0.9 R_J$ leading to a
density of $2.21\rho_J$. The reported semi-major axis is $0.0142$ AU.

In order to improve the characterization of \wq and its host star, for
which \cite{husnoo12} find evidences of excess rotation,
\cite{gillon12} observed over thirty eclipses between transits and
occultations, using mainly the TRAPPIST telescope in the $I+z$ and $z$
bands, respectively. The survey was completed with transit
observations obtained with the Euler telescope in the Gunn $r'$ band
and with the VLT/HAWK-I instrument installed on the $8.2\meter$ UT4
telescope in Paranal.  The work allowed significant improvements in
the determination of the stellar density and in the refinement of the
parameters of the planetary system.  In particular,
\cite{gillon12} find a mass of $2.034 M_J$ and a radius of
$1.036 R_J$. Occultations also detected the thermal emission of the
planet.

Thermal emission is also the subject of Spitzer observations, carried
out by \cite{blecic12, blecic14}, which highlighted the possibility of
thermal inversion in the atmosphere. 

\cite{line14} uniformly analyzed occultations of nine exoplanets
including \wq, with the goal to characterize their $\rm C$ to $\rm O$
ratio, which is found to be $>1$ in the case of \wq, and which is
attributed to the upper limit value of $\rm CH_4$. 

Other observations using the GROND instrument on the MPG/ESO
$2.2\meter$ telescope in the $g'$, $r'$, $i'$, $z'$, $J$, $H$, and $K$
bands \citep{chen14} allowed to refine the period to a
value of $P=0.81347437$. This value agrees with spectroscopic
observations carried on by \cite{murgas14}. These
observations also confirmed planetary dayside thermal emission with
poor heat redistribution, as observed by \cite{wang13,
  wang13-proc}. 

\begin{table}[t]
  \caption{Log of the observations of \ww and \wq. \label{tab:log}}
\centering
\begin{tabular}{cccc}
\tableline
\tableline
    Date    & Filter &        Exp.         & Telescope    \\
 \tableline                          
 & \multicolumn{3}{c}{\ww} \\ \cline{2-4}
 2014-03-17 &  $U$   & $240$--$300\second$ & $2.12\meter$  \\
 2014-03-17 &  $I$   & $120$--$180\second$ & $0.84\meter$  \\
 2014-03-21 &  $R$   &    $120\second$     & $0.84\meter$  \\
 \tableline
 & \multicolumn{3}{c}{\wq} \\ \cline{2-4}
 2014-02-13 &  $R$   & $60$--$120\second$  & $0.84\meter$  \\
 2014-03-07 &  $i$   &    $140\second$     & $1.50\meter$  \\
 2014-03-07 &  $R$   &     $60\second$     & $0.84\meter$  \\
 2014-03-20 &  $I$   & $60$--$120\second$  & $0.84\meter$  \\
 2014-03-21 &  $V$   & $60$--$120\second$  & $0.84\meter$  \\
 2014-03-29 &  $i$   &     $90\second$     & $1.50\meter$  \\
 2014-05-12 &  $I$   & $60$--$120\second$  & $0.84\meter$  \\
\tableline
\end{tabular}
\end{table}

Finally, an X-ray study by \cite{czesla13} on the host star of \wq
using XMM-Newton observations found that the planet is exposed to a
high-energy radiation field; 
and water abundance measurements were recently reported by
\cite{laura14}.

In Sect.~\ref{obs} we describe the used telescopes and our
observations of \ww and \wq, while in Sect.~\ref{data}
and~\ref{sect:p} we focus on data reduction techniques and on a
discussion about the period of the two objects.  Results are presented
in Sect.~\ref{disc}, while conclusions are described in
Sect.~\ref{conc}.

\section{Observations}
\label{obs}


Observations were carried on using all three equatorial
Ritchey-Chr\'etien telescopes installed at the the San Pedro M\'artir
National Astronomical Observatory (SPM-OAN). These facilities are
located at an altitude of $2\,800\meter$ in the middle of the
peninsula of Baja California in northwestern Mexico (N~\ang{31;02;39},
W~\ang{115;27;49}). The $0.84\meter$ and the $2.12\meter$ are operated
on-site, and the $1.50\meter$ is a robotic telescope monitored and
operated remotely.

We principally used the $0.84\meter$
telescope, 
an $f/15$ which provides an $8.4\arcmin\times19.0\arcmin$ field of
view on the \emph{Mexman} instrument.  We decided to window the
\emph{Esopo}
CCD 
to avoid vignetting and to reduce the readout time. The instrument
comes with a set of standard $UBVRI$ Johnson filters among others.
%
%
In this paper we present observations carried out using the $VRI$
filters.

\begin{figure}[t]
  \centering
  \includegraphics[width=\columnwidth]{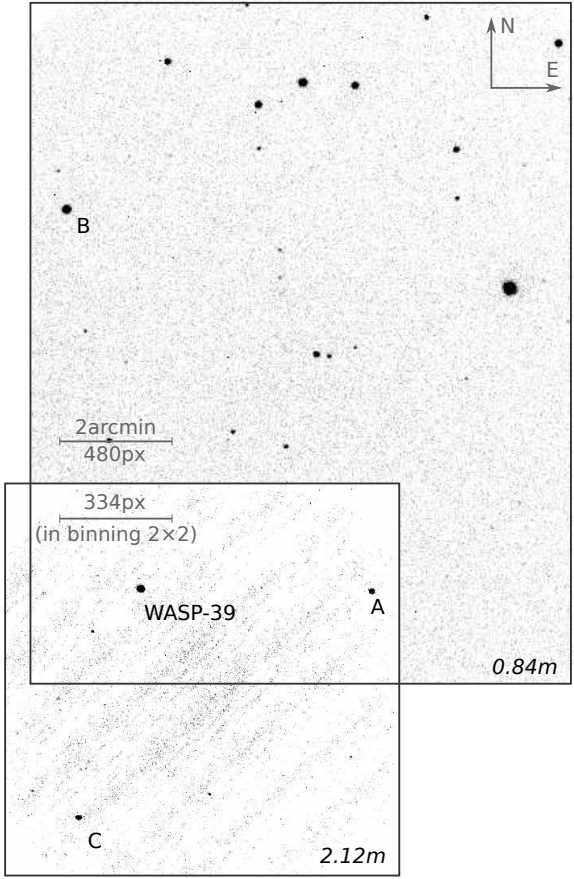}
  \caption{Superposed fields of \ww obtained with the $0.84\meter$ and
    the $2.12\meter$ telescopes, in two random CCD images, and using
    an inverse color scale. The target and  field stars A, B, and C
    are also shown. Star A was chosen as a reference for photometry
    for both the telescopes. \label{w39-field}}
\end{figure}

\begin{figure}[t]
  \centering
  \includegraphics[width=\columnwidth]{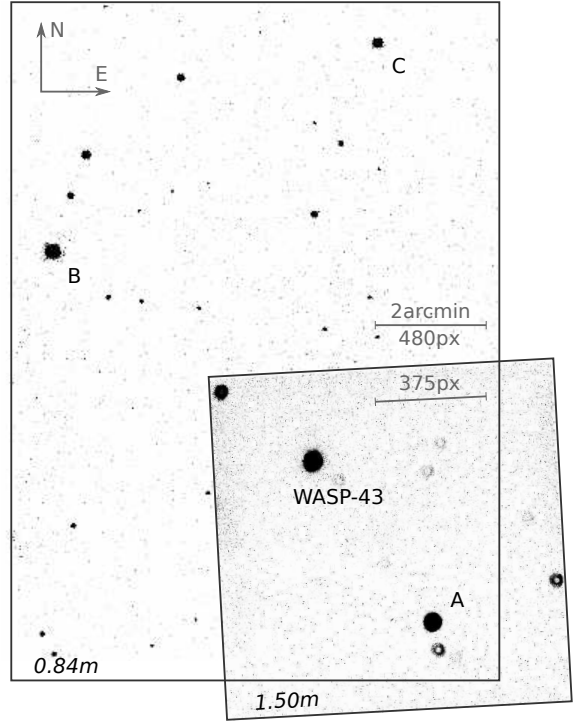}
  \caption{Superposed fields of \wq as obtained with the $0.84\meter$
    and the $1.50\meter$ telescopes, in two random CCD images, and
    using an inverse color scale. The defocusing is particularly
    appreciable in the $1.50\meter$ telescope images. The target field
    stars A, B, and C are also shown. Star A was chosen as a reference
    for photometry for both the telescopes. \label{w43-field}}
  
\end{figure}

\notetoeditor{Figures \ref{w39-field} and \ref{w43-field} should be
  placed side by side on top on the same page}

Observations were also obtained with the robotic $1.50\meter$
telescope 
and its multi-channel imager \emph{RATIR} \citep{2011AAS...21715707R,
  2012SPIE.8444E..5LW, 2012SPIE.8446E..10B, 2012SPIE.8446E..9OF}.
This instrument, designed for Gamma Ray Burst observations and
follow-up, is equipped with three dichroics to transmit the light to
four detectors named C0, C1, C2, and C3, equipped with optical and
infrared filters.



The CCDs work in $2\times 2$ binning mode
by default.  In this paper we present observations carried out using
the Gunn-$i$ only, corresponding to the C1 detector which provides a
$5.6\arcmin\times5.6\arcmin$ field of view.

Taking advantage from free observing time during maintenance
operations, we had the possibility to use the $2.12\meter$
telescope, 
improved with active optics on the primary mirror, which is normally
dedicated to spectroscopy.  We used this telescope in direct-imaging
mode.  The observation presented in
this paper was carried out using the $U$ filter on the \emph{Marconi}
CCD 
providing a $6.2\arcmin\times6.2\arcmin$ field of view.

Additional details about telescopes and detectors are listed in
Table~\ref{tab:tel}, while Table~\ref{tab:log} shows the log of the
observations.

\begin{figure}[t]
  \centering
  \includegraphics[width=\columnwidth]{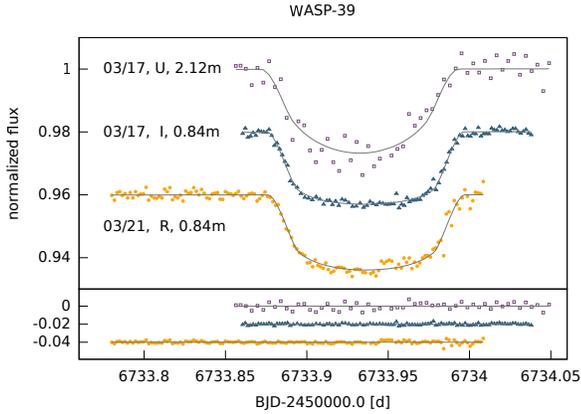}
  \caption{Light curves of \ww obtained with the $0.84\meter$
    telescope in $R$ and $I$ bands and with the $2.12\meter$ telescope
    in the $U$ band. For displaying purposes, the light curves were reported
    to the time of the last observation (2014-03-21), and an
    arbitrarily shift is applied.  The fit of the light curves (solid
    lines) is also shown, as well as the residuals in the lower box.
    \label{w39-curves}}
\end{figure}

\begin{figure}[t]
  \centering
  \includegraphics[width=\columnwidth]{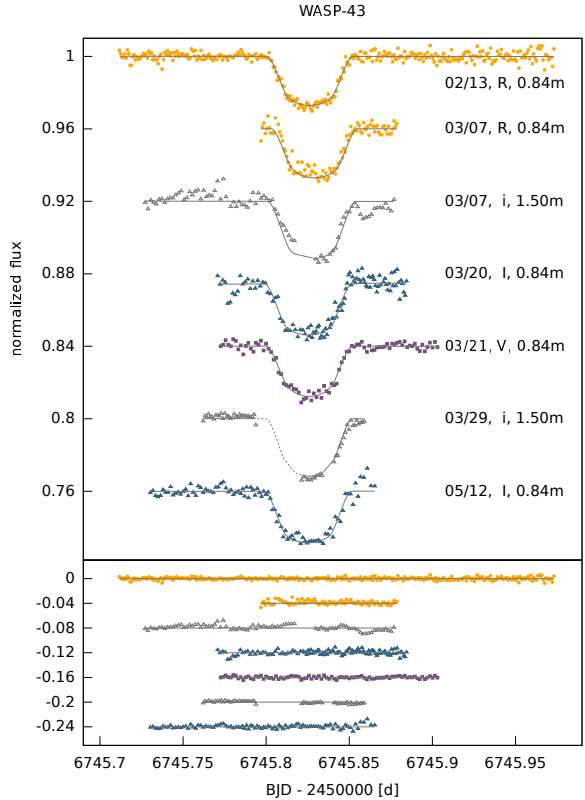}
  \caption{Light curves of \wq obtained with the $0.84\meter$
    telescope in $VRI$ bands and with the $1.50\meter$ telescope in
    the Gunn-$i$ band, reported to the time of the observation of 
    2014-03-29, and arbitrarily shifted. The fit and the residuals 
    are also shown. The dashed line represent an interpolation of the 
    model.
    \label{w43-curves} }
\end{figure}

\notetoeditor{Figures \ref{w39-curves} and \ref{w43-curves} should be
  placed side by side on top on the same page}

\subsection{\ww}
\label{obs:ww}

We performed three observations of \ww, covering two subsequent
transits.  Its parent star is a G-type dwarf with a reported magnitude
of $V=12.1$ \citep{w39}.

For the transit of 2014-03-17 we had simultaneous access to the
$0.84\meter$ and to the $2.12\meter$ telescopes.  Fig.~\ref{w39-field}
shows the target and its nearby stars as seen in the fields of the two
telescopes, some of which (A, B and C) were tested as comparison
stars, as seen by the fields of the two telescopes.  We observed in
the $I$ band with the smaller telescope, obtaining 103 frames with
exposure times of $120\second$ and $180\second$, and in the $U$ band
with the larger one, in order to take advantage from its sensitivity,
obtaining 57 frames with exposure times of $240\second$ and
$300\second$.  As the object was particularly faint in this filter, we
decided to use a $2\times2$ binning and a small defocus.

The transit of 2014-03-21 was observed with the $0.84\meter$ telescope
only.  We obtained 142 frames in the $R$ band with an exposure time of
$120\second$, slightly defocusing the telescope to avoid saturation.

The night conditions during all observations suffered from very
thin high haze and a nearly-full Moon, with a seeing of
about~$3$--$4\arcsec$.
Depending on the instrument, the band, and seeing conditions,
different defocusing levels were used.  Moreover, rapid changes in
the atmospheric conditions required a variation of the exposure time
in order to avoid saturation.

\subsection{\wq}
\label{obs:wq}

Due to its short period with respect to \ww, we had the opportunity to
carry out more observations of \wq.  We obtained a total of seven
transits spanning six nights.  Fig.~\ref{w43-field} shows two fields
containing the target and nearby stars as seen with the $0.84\meter$
and $1.50\meter$ telescopes.

Four transits were observed with the $0.84\meter$ telescope only: on
2014-02-13, 2014-03-20, 2014-03-21, 2014-05-12 in the $R$, $I$, $V$,
and again $I$ filters, respectively, with exposure times spanning from
$60\second$ to $120\second$.

Another transit was obtained with the $1.50\meter$ telescope only
(2014-03-29), strongly defocused, in the Gunn-$i$ filter, and with a
$90\second$ exposure time. This transit was unfortunately affected by
clouds from the first contact until almost the mid time.

Finally, we observed with both the $0.84\meter$ and $1.50\meter$
telescopes on 2014-03-07, using the $R$ and Gunn-$i$ filters
respectively, and with exposure times of $60\second$ and $140\second$.

All observations were taken in condition of a nearly-full moon.

\section{Data reduction}
\label{data}

CCD images of both transits were debiased and flat-fielded using
standard IRAF procedures from the \textsc{ccdproc} package.  Master
bias and master sky flat fields were obtained from images taken at the
beginning and at the end of each observing night.  Cosmic rays were
also removed using the method implemented by the \textsc{lakos}
package \cite{2001PASP..113.1420V}, then images were aligned using the
\textsc{imalign} IRAF procedure.

We decided to use different aperture photometry routines to obtain the
light curves: a custom IRAF routine and the \texttt{defot} routine
\citep{2010MNRAS.408.1680S} written in the IDL language, that we
modified to work with the FITS headers generated by the SPM-OAN
telescopes.  We found that the results of these two pipelines are in
good agreement.  In the case of images with a small amount of defocus,
it was possible to calculate the Full Width Half Maximum (FWHM) of the
Point Spread Function by fitting an elliptical Gaussian.  We used this
information to dynamically adapt the aperture radius to a value of 2.5
times the calculated FWHM.  We found that this technique increases the
quality of the photometry with respect to the fixed-radius aperture.
In the case of images with a consistent defocus, several fixed
apertures were tested in order to find the optimal value.

Several field stars were tested in order to find a reference star to
perform differential photometry, and a non-saturated star with a color
index and magnitude values as close as possible to those of the target
star was chosen. The chosen reference stars for \ww and \wq are
labeled as ``A'' in Fig.~\ref{w39-field} and Fig.~\ref{w43-field},
respectively.

We then used the target and the reference star to produce differential
light curves.

The resulting light curves may present some residual trends, which were
removed with a first-order airmass correction described by
\cite{fox}.

The timestamps of the light curve were converted to the Dynamical
Time-based system ($BJD_{\rm TDB}$, here therefore $BJD$) using the
transformation given by \cite{eastman10}.

We used the aperture photometry technique presented above for both
\ww, whose three final light curves are presented in
Fig.~\ref{w39-curves}, and \wq, whose seven final light curves are
shown in Fig.~\ref{w43-curves}.

\subsection{Fit}

\begin{table*}[t]
  \caption{Fit results of the physical and orbital 
    parameters of \ww and \wq obtained with TAP. 
    The fit of the period and the period variation are described 
    in Sect.~\ref{sect:p}. \label{params}}
  \quad\centering
  \begin{tabular}{lccc}
\tableline
\tableline
    Parameter      & Filter &      WASP-39      & WASP-43 \\
\tableline             
  $i$ [\degree]    &        &  $87.78\pm0.43$   & $81.92\pm0.54$ \\
     $a/R_*$       &        &  $11.32\pm0.42$   & $4.82\pm0.11$ \vspace{0.15cm} \\
                   &  $U$   & $0.1462\pm0.0116$ & \\
                   &  $V$   &                   & $0.1615\pm0.0041$ \\
    $R_p/R_*$      &  $R$   & $0.1424\pm0.0023$ & $0.1599\pm0.0025$ \\
                   &  $I$   & $0.1424\pm0.0023$ & $0.1653\pm0.0054$ \\
                   &  $i$   &                   & $0.1738\pm0.0033$ \\
\tableline
     $P$ [d]       &        &     4.055259      & 0.81347437 \\
       $e$         &        &         0         & 0 \\
$\omega$ [\degree] &        &         0         & 0 \\
                   &  $U$   &       0.950       & \\
                   &  $V$   &                   & 0.750 \\
      $l_1$        &  $R$   &       0.425       & 0.599 \\
                   &  $I$   &       0.335       & 0.451  \\
                   &  $i$   &                   & 0.485 \vspace{0.15cm}\\
                   &  $U$   &      -0.086       & \\
                   &  $V$   &                   & 0.040 \\
      $l_2$        &  $R$   &       0.246       & 0.137 \\
                   &  $I$   &       0.250       & 0.193 \\
                   &  $i$   &                   & 0.183 \\
\tableline
\end{tabular}
\tablecomments{
  The upper part of the table shows the fitted             
  parameters, while the lower part shows the fixed values: 
  $P$ from \cite{w39} for \ww and from                     
  \cite{chen14} for \wq; $l_1$ and $l_2$ from the          
  Exofast online tool \citep{eastman13}.
}
\end{table*}

The light curves were fitted using the IDL software Transit Analysis
Package (TAP) implemented by \cite{gazak11}, which uses a Markov Chain
Monte Carlo (MCMC) method to find the best fit parameters for the
\cite{mandel02} model.

This code allows to fit multiple light curves simultaneously, which is
particularly useful for fixing parameters such as the orbital
inclination $i$ and the scaled semi-major axis $a/R_*$ to the same
values for a group of curves, and thus obtain a global fit for these
quantities.

Other parameters, such as the mid-transit time $T_{mid}$ and the
scaled planetary radius $R_p/R_*$ are allowed to be fitted
individually for each curve.  This technique allows to fit the same
orbital model to a group of light curves of the same object obtained
in different bands, and to find potential planetary radius differences
as a function of the wavelength.
 
All light curves of each system were fitted simultaneously.
We used a set of fixed values in the MCMC analysis for several
parameters: the Period $P$, taken from \cite{w39} for \ww and from
\cite{chen14} for \wq; the eccentricity $e$ and the argument of
periastron $\omega$, both set $=0$.
TAP allows the use of linear and quadratic models for the stellar limb
darkening. A quadratic model is assumed in this analysis, and the two
terms $l_1$ (linear) and $l_2$ (quadratic) are fixed in the MCMC
analysis to the values corresponding to the filter used for each light
curve.  The values were obtained from the \textsc{exofast}
\citep{eastman12,eastman13} online
tool\footnote{\url{http://astroutils.astronomy.ohio-state.edu/exofast/limbdark.shtml}},
which are interpolated from stellar atmosphere models of
\cite{claret11}.

We then fitted for each system: the orbital inclination $i$, the
scaled semi-major axis $a/R_*$, and a value of the scaled radius
$R_p/R_*$ for each one of the used filters. TAP was initialized with
the most recent parameters reported in \cite{w39} for \ww and
\cite{chen14} for \wq.

The mid-transit time $T_{mid}$ for each light curve was also fitted
(in the case of multiple light curves of the same object obtained on the
same night using different telescopes, the same $T_{mid}$ was fitted
for both the light curves). For the MCMC analysis, the $T_{mid}$ is initialized
with a value estimated by TAP from the input light curve.

The best-fit models and corresponding residuals are shown in
Figs.~\ref{w39-curves}~and~\ref{w43-curves}, and the values of the
parameters are shown in Table~\ref{params}.

\section{Period and period variation}
\label{sect:p}

Using the fitted values for the mid-transit time $T_{mid}$, we
retrieved the period $P$ of \ww and \wq by fitting the following
linear law:
\begin{equation}
  T_{mid} = T_0 + NP
  ,
  \label{eq:p}
\end{equation}
where $T_0$ is the initial epoch and $N$ is the number of periods
since $T_0$.  $T_0$ and $P$ are best-fit parameters.

\begin{figure}[t]
  \centering
  \includegraphics[width=\columnwidth]{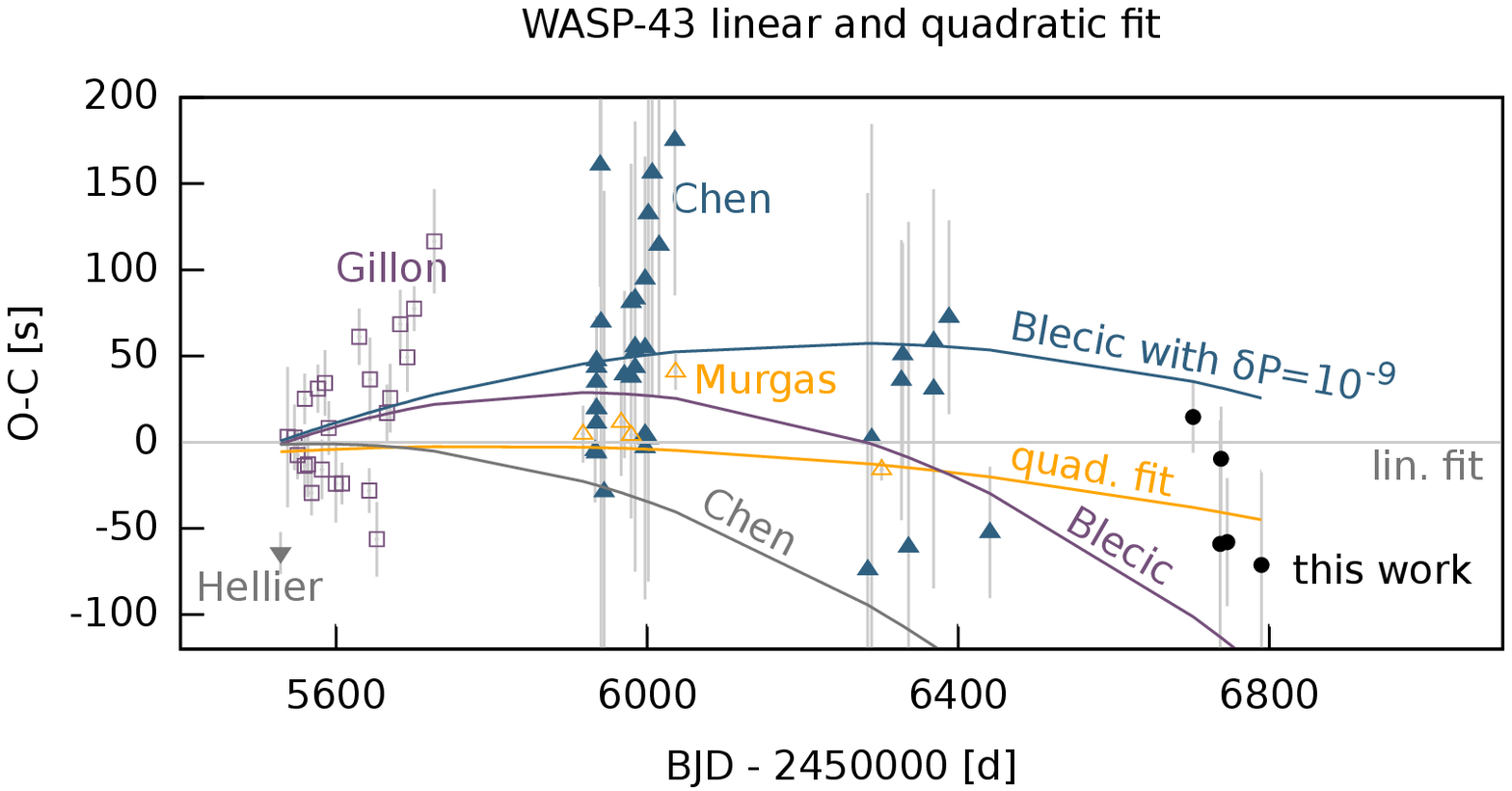}
  \caption{Observed $-$ calculated $T_{mid}$, linear and quadratic fit
    of the \wq period including data from
    \cite{w43,gillon12,chen14,murgas14}, and from our
    observations. See Sect.~\ref{w43-quad-discussion} for curve
    details. \label{w43-quad}}
\end{figure}

\notetoeditor{Figures \ref{w39-period} and \ref{w43-period} should be
  placed side by side on top on the same page; Figure \ref{w43-quad}
  should be placed under Figure \ref{w43-period} }

\begin{figure}[t]
  \centering
  \includegraphics[width=\columnwidth]{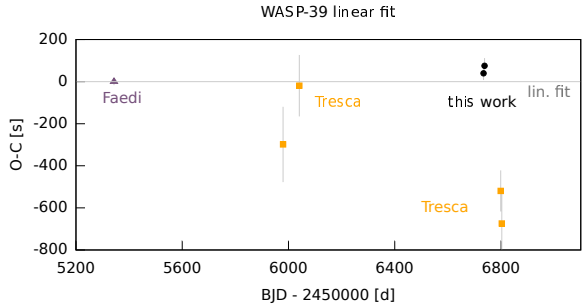}
  \caption{Observed $-$ calculated $T_{mid}$ and linear fit of the
     \ww period, including data from \cite{w39}, from
     the ETD-TRESCA database, and from our observations. 
     \label{w39-lin}}
 \end{figure}

  
 
 Further analysis of the ephemerids was carried out in order to
 investigate a long-term variation of the period with respect to the
 time. A constant decrease of the period for close-orbiting planets
 may be indicative of processes that remove orbital energy such as
 tidal dissipation \citep{adams10a,adams10b,sasselov03,levrard09}.
 A simple model assuming a constant variation of the period is given
 by the following quadratic equation
 \begin{equation}
   T_{mid} = T_0 + NP + \delta P N(N-1)/2
   \label{eq:dp}
 \end{equation}
proposed by \cite{adams10b}, where $\delta P = P\dot{P}$.


For the present analysis, a Levenberg-Marquardt least-squares fitting
algorithm implemented in the code of \cite{mark09} was initially used
to find the best-fit parameters of Eqs.~\ref{eq:p} and~\ref{eq:dp}.
However, this algorithm can be trapped in a local minimum of
$\chi^2$. For this reason, a Monte Carlo code was set-up to search for
the $\chi^2$ minimum and sample the parameter space. Both approaches
were tested on the \object{OGLE-TR-113b} data provided by
\cite{adams10a}, finding good agreement with their results.


We find that both approaches are consistent and give similar best-fit
values.  Comforted by these results, we decided to apply this method
to our data, and we describe the results in the following subsections.

\subsection{\ww}

Concerning \ww, the $T_{mid}$ reported by \cite{w39} and the
  one calculated from our observations are fitted using
  Eq.~\ref{eq:p}. This gives a
\begin{itemize}
\item $T_0 = 2455342.9687998 \pm 0.0001999$ (BJD), 
\item $P = 4.0552989 \pm 9.963 \times 10^{-7}$ days. This fit has a
\item $\chi^2 = 0.541$,
\end{itemize}
where $\chi^2$ stands for the ``reduced $\chi^2$'', i.e. $\chi^2/D_f$,
where $D_f$ are the degrees of freedom (number of data $-$ number of
parameters).

However, due to the significantly small number of observations, we
decided to increase the number of transits considered in our analysis
by introducing additional data from the Exoplanet Transit Database
(ETD)
webpage 
\citep{poddany10}, in order to improve the robustness of our results.
We selected a total of three light curves. According to this database,
the timings of the light curves are reported in HJD. They were
transformed to BJD using the software provided by \cite{eastman10}. We
used TAP to obtain only $T_{mid}$ for these light curves by running
the MCMC analysis with all other parameters to be fixed to the
values provided by \cite{w39}. By adding these new data to previous
ones, we obtain a total of 7 observations.
The linear fit of Eq.~\ref{eq:p} was repeated with these additional
transits, and the following results were obtained:
\begin{itemize}
\item $T_0 = 2455342.9695727 \pm 0.000199$ (BJD),
\item $P = 4.0552947 \pm 9.651 \times 10^{-7}$ days, and a
\item $\chi^2 = 11.82$,
\end{itemize}
indicating a period larger than that reported by \cite{w39}. The
residuals for this fit are shown in Fig.~\ref{w39-lin}.
As the last two Tresca points show large differences with
  respect to the fit, we tested the fit process without their
  contribution, obtaining the following results:
\begin{itemize}
\item $T_0 = 2455342.9695554 \pm 0.0001994$ (BJD),
\item $P = 4.0552965 \pm 9.956 \times 10^{-7}$ days, and a
\item $\chi^2 = 1.299$,
\end{itemize}
and in this case too we report a larger period with respect to
  \cite{w39} results.  In both cases the results are consistent, but
  we suggest to improve their robustness with additional
  observations.

An attempt to fit Eq.~\ref{eq:dp} was made, but the small amount of
data do not allow a reasonable result for the fit.
More data are needed as well as precise timings, in order to assess if
there is a long-term constant variation of the period, or to detect
the presence of other transit timing variations.


\subsection{\wq}
\label{w43-quad-discussion}

A sufficient amount of \wq transit timings is available in the
literature spanning several years of observations
\citep{w43,gillon12,chen14,murgas14}. 

Some previous studies of possible long-term variations have been
carried out by \cite{chen14} and \cite{murgas14}, suggesting that the
period of \wq is slowly decreasing with time. In this work, we repeat
the analysis combining our data with the timings available in the
literature. A best-fit model was obtained using the
Marquardt-Levenberg algorithm previously described. For the linear
fit, best-fit values are
\begin{itemize}
\item $T_0 = 2455528.86839966 \pm 0.000036904078$ (BJD), and
\item $P = 0.81347409 \pm 7.3\times 10^{-8}$ days, giving a
\item $\chi^2 = 3.784$.
\end{itemize}
For the quadratic fit, best fit values are 
\begin{itemize}
\item $T_0 = 2455528.86848347 \pm 0.0000369043$ (BJD),
\item $P = 0.81347374 \pm 7.285 \times 10^{-8}$ days, and a
\item $\delta P = 6.7 \pm 6.9 \times 10^{-10}$, corresponding to 
\item $\dot{P} = 0.03 \pm 0.03\second/\rm{year}$ .  This gives a
\item $\chi^2 = 3.819$.
\end{itemize}
The small difference in $\chi^2$ between the linear and quadratic fits
shows that, statistically, there is no significant improvement by
using the latter, as shown in Fig.~\ref{w43-quad}.

Using the values provided by \cite{blecic14}, we obtain
\begin{itemize}
\item $P = 0.81347530 \pm 3.9 \times 10^{-7}$ days, and
\item $\dot{P} = -0.095 \pm 0.036\second/\rm{year}$, giving a
\item $\chi^2 = 21.25$.
\end{itemize}
The parameters reported by \cite{chen14} are
\begin{itemize}
\item $P = 0.81347399 \pm 2.2\times 10^{-7}$ days, and we have
\item $\dot{P} = -0.09 \pm 0.04\second/\rm{year}$, giving a
\item $\chi^2 = 18.26$.
\end{itemize}
The fit was also tested with the Monte Carlo approach which find a a
minimum $\chi^2$ corresponding to $P = 0.81347382$ and $\dot{P} =
0.02\second/\rm{year}$, with a $\chi^2 = 4.046$. This analysis showed
that the $\chi^2$ minimum is very close to a value of $\dot{P} = 0$.

These fits are visualized in Fig.~\ref{w43-quad}. The model of
\cite{blecic14} fits most of the points reasonably well, except for
those corresponding to our observations.  The model of \cite{chen14}
shows more deviations with respect to recent observations. For
comparison, we also plot a third model using the same period reported
by \cite{blecic14} but with a
\begin{itemize}
\item $\delta P = -1.0 \times 10^{-9}$, which is in the order of
  magnitude of previously reported values, giving a
\item $\chi^2 = 46.96$. 
\end{itemize}
This corresponds to the same orbital period of
\cite{blecic14} but with a smaller value of $\dot{P} =
-0.038\second/\rm{year}$. It can be seen that it fits reasonably
points from earlier observations. 

However, a detailed analysis of Transit Time Variations has not been
considered in this work, which could explain some of the observed
variations \citep{chen14}.

The present analysis then confirms the results of \cite{chen14}, also
showing that a quadratic model does not improve significantly the fit
of the transit timings for \wq.

\section{Results and discussion}
\label{disc}



We report for the first time a light curve of WASP-39b in the $U$ band
obtained with the $2.12\meter$ telescope at SPM-OAN.  Observations in this
band are important to study the presence of high-hazes in the
atmosphere, as Rayleigh scattering becomes important at smaller
wavelengths, which can produce larger observed radii in this band
\citep{2012MNRAS.422.3099S,2013MNRAS.434..661C}. The resulting light
curve is shown in Fig.~\ref{w39-curves}. Although there is more
dispersion in the data compared to the $I$ and $R$ bands, a depth flux
of approximately $3\%$ is observed. The best-fit values obtained in
the previous section give a relative radius of $R_p/R_* =
0.1462 \pm 0.0116$.

No significant variations of the planetary radius between the $U$,
$R$, and $I$ bands is found. It is also noted that the $U$ and $I$
band light curves were obtained in the same night, and no evident
asymmetries are observed that could suggest the presence of tails or
other features around the planet. Additional observations in the $U$
band can help to reduce the uncertainty for this highly inflated
extrasolar planet.

The analysis of the ephemerids in the previous section gives a period
for \ww approximately $3.084\pm0.774\second$ larger than that of
\cite{w39}.  

\begin{figure}[t]
  \centering
  \includegraphics[width=\columnwidth]{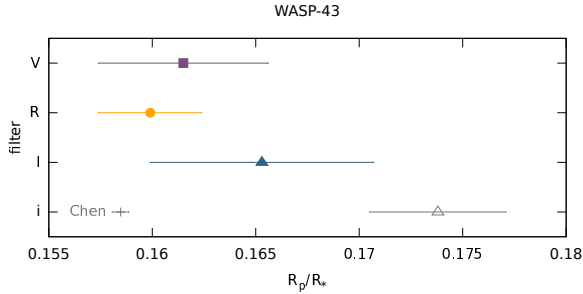}
   \caption{\wq planet radius in the observing filters (see
    Table~\ref{params}).
    The value reported by \cite{chen14}
    in the $i$ filter is also shown.
  \label{w43-radius}}
\end{figure}

\wq shows a slight dependence of the $R_p/R_*$ as a function of the
used filter (see Fig.~\ref{w43-radius}), which should be investigated
with additional observations.  
In particular, we find for filter $i$ a
value higher by  $0.01637\pm0.00371$ with
respect to the value reported by \cite{chen14} in the same
filter. ($0.1738\pm0.0033$ against $0.15847\pm0.00041$).
Although, as our two $i$ filter curves are not complete and show
signatures of red noise, we suggest to carry on more observations to
confirm the significancy of this result.

Nevertheless, we find a tendency for the $R_p /R_*$ similar to that of
\cite{chen14} in the sense that the planetary radius is significantly
higher at the $i$ band with respect to the value in other bands. 
The semi-major axis is slightly smaller, but comparable
to the value obtained by \cite{gillon12} and \cite{chen14}.

The analysis of the ephemerids gives a period consistent with
previous results, and we confirm that there is no improvement while
using a quadratic model for fitting the ephemerids. Further analysis
and observations would be required to ascertain if a constant period
variation exists for \wq. Given the proximity of this planet to
its host star, this is particularly relevant to explore potential
mechanisms that remove orbital energy such as tidal dissipation.\\

The present work is part of an ongoing survey started in 2014. A total
of 23 extrasolar planets were observed mainly with the $0.84\meter$
and the $1.50\meter$ telescopes giving a total of 40 good quality
light curves in several filters. Upgrades in the instrumentation, refinement
of the observing technique, and the possibility to benefit from dark
and photometric nights will increase the quality of the results in the
near future, and will complement other telescope data for multi-site
studies and analysis of this class of objects.
%
%
Moreover, three new $1.30\meter$ telescopes will be completed by 2016
at the San Pedro M\'artir observatory, which are part of the TAOS II
project \citep{lehner-spie, geary12, lehner13}.
%

The characteristics of this survey \citep{lehner12} allow to set up
specific pipelines for the early detection of new extrasolar planets,
for example with a binning of the light curves as already tested for
TAOS data \citep{2014arXiv1408.6189R}.
%
%
Also, a project for the installation of a $6.5\meter$ telescope at
the San Pedro M\'artir observatory has recently reached the phase of
preliminary design.

Current and \emph{in fieri} SPM-OAN telescopes then represent a
unique opportunity for the extrasolar planet community to complement
the northern hemisphere observatories involved in this research area.

\section{Conclusions}
\label{conc}

Multi-filter exoplanet transit observations of \ww and \wq, carried on
for the first time with all three San Pedro M\'artir telescopes, have
shown the capability of these facilities of $0.84\meter$, $1.50\meter$
and $2.12\meter$ for this kind of investigation.  The fit of the two
objects shows a scatter of $1.5$--$2.5\milli\rm mag$ rms in condition
of full or nearly-full Moon. This makes these instruments suitable for
future ground-based observing campaigns, in particular for the
follow-up of alerts triggered by the upcoming projects currently in
development.

The analysis shows most of the parameters to be in good agreement with
previous works, in particular for what concerns the first observation
\ww in the $U$ filter.

The period for \ww is found to be significantly larger ($>3\second$)
with a $4\sigma$ accuracy with respect to previous works.

The value of the planet/star radius of \wq is  larger in
the $i$ filter with respect to previous works.  

A tendency of a higher planet radius in the $i$ band is obtained,
consistent with the tendency reported in recent works.  Additional
observations of this object are needed to confirm a slight dependence
of this parameter on the observing filter.

We suggest to plan additional observations in order to confirm
  the accuracy of these result with respect to the literature. 

\acknowledgments

Research carried out thanks to the support of UNAM-DGAPA-PAPIIT
project IN115413.  FGRF acknowledges the support of a CONACYT
scholarship for postgraduate studies in Mexico.  We also acknowledge
Dr. Wolfgang Steffen Burg for the discussion about future
exoplanetary-related projects, John Southworth for his useful comments
about the form ant the content of this manuscript, and the OAN
technical and service staff for daily support.  We finally acknowledge
the anonymous referee for the useful remarks.

{\it Facilities:} 
\facility{San Pedro M\'artir $0.84\meter$ (MEXMAN)}, \\
\facility{San Pedro M\'artir $1.50\meter$ (RATIR)}, \\
\facility{San Pedro M\'artir $2.12\meter$ (direct imaging mode)}. 



\end{document}